# UNET 3+: A FULL-SCALE CONNECTED UNET FOR MEDICAL IMAGE SEGMENTATION

*Huimin Huang[1], \*Lanfen Lin[1], Ruofeng Tong[1], \*Hongjie Hu[2], Qiaowei Zhang[2], Yutaro Iwamoto[3], Xianhua Han[3], \*Yen-Wei Chen[3,4,1], Jian Wu[1]*

[1] College of Computer Science and Technology, Zhejiang University, China
[2] Department of Radiology, Sir Run Run Shaw Hospital, China
[3] College of Information Science and Engineering, Ritsumeikan University, Japan
[4] Research Center for Healthcare Data Science, Zhejiang Lab, Hangzhou, China

\*Corresponding Authors: Lanfen Lin (llf@zju.edu.cn), Hongjie Hu (hongjiehu@zju.edu.cn), Yen-Wei Chen (chen@is.ritsumei.ac.jp)

## ABSTRACT

Recently, a growing interest has been seen in deep learning-based semantic segmentation. UNet, which is one of deep learning networks with an encoder-decoder architecture, is widely used in medical image segmentation. Combining multi-scale features is one of important factors for accurate segmentation. UNet++ was developed as a modified Unet by designing an architecture with nested and dense skip connections. However, it does not explore sufficient information from full scales and there is still a large room for improvement. In this paper, we propose a novel UNet 3+, which takes advantage of full-scale skip connections and deep supervisions. The full-scale skip connections incorporate low-level details with high-level semantics from feature maps in different scales; while the deep supervision learns hierarchical representations from the full-scale aggregated feature maps. The proposed method is especially benefiting for organs that appear at varying scales. In addition to accuracy improvements, the proposed UNet 3+ can reduce the network parameters to improve the computation efficiency. We further propose a hybrid loss function and devise a classification-guided module to enhance the organ boundary and reduce the over-segmentation in a non-organ image, yielding more accurate segmentation results. The effectiveness of the proposed method is demonstrated on two datasets. The code is available at: *github.com/ZJUGiveLab/UNet-Version*

## 1. INTRODUCTION

Automatic organ segmentation in medical images is a critical step in many clinical applications. Recently, convolutional neural networks (CNNs) greatly promoted to developed a variety of segmentation models, e.g. fully convolutional neural networks (FCNs) [1], UNet [2], PSPNet [3] and a series of DeepLab version [4-6]. Especially, UNet, which is based on an encoder-decoder architecture, is widely used in medical image segmentation. It uses skip connections to combine the high-level semantic feature maps from the decoder and corresponding low-level detailed feature maps from the encoder. To recede the fusion of semantically dissimilar feature from plain skip connections in UNet, UNet++ [7] further strengthened these connections by introducing nested and dense skip connections, aiming at reducing the semantic gap between the encoder and decoder. Despite achieving good performance, this type of approach is still incapable of exploring sufficient information from full scales.

As witnessed in many segmentation studies [1-7], feature maps in different scale explore distinctive information. Low-level detailed feature maps capture rich spatial information, which highlight the boundaries of organs; while high-level semantic feature maps embody position information, which locate where the organs are. Nevertheless, these exquisite signals may be gradually diluted when progressively down- and up-sampling. To make full use of the multi-scale features, we propose a novel U-shape-based architecture, named UNet 3+, in which we re-design the inter-connection between the encoder and the decoder as well as the intra-connection between the decoders to capture fine-grained details and coarse-grained semantics from full scales. To further learn hierarchical representations from the full-scale aggregated feature maps, each side output is connected with a hybrid loss function, which contributes to accurate segmentation especially for organs that appear at varying scales in the medical image volume. In addition to accuracy improvements, we also show that the proposed UNet 3+ can reduce the network parameters to improve the computation efficiency.

To address the demand for more accurate segmentation in medical image, we further investigate how to effectively reduce the false positives in non-organ images. Existing methods solve the problem by introducing attention mechanisms [8] or conducting a pre-defined refinement approach such as CRF [4] at inference. Different from these methods, we extend a classification task to predict the input image whether has organ, providing a guidance to the segmentation task.

In summary, our main contributions are four-fold: (i) devising a novel UNet 3+ to make full use of the multi-scale features by introducing full-scale skip connections, which incorporate low-level details with high-level semantics from feature maps in full scales, but with fewer parameters; (ii) developing a deep supervision to learn hierarchical representations from the full-scale aggregated feature maps, which optimizes a hybrid loss function to enhance the organ boundary; (iii) proposing a classification-guided module to reduce over-segmentation on none-organ image by jointly training with an image-level classification; (iv) conducting extensive experiments on liver and spleen datasets, where UNet 3+ yields consistent improvements over a number of baselines.

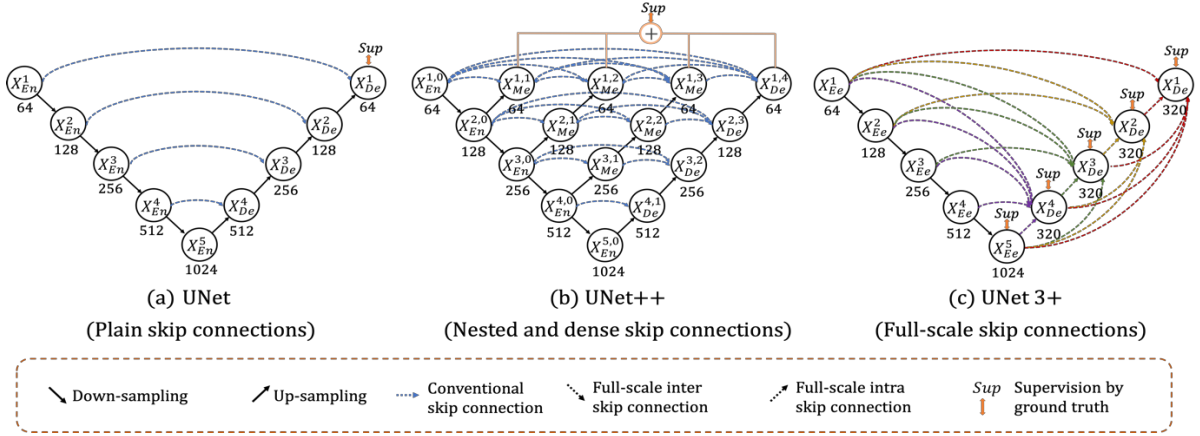

Fig .1: Comparison of UNet (a), UNet++(b) and proposed UNet 3+ (c). The depth of each node is presented below the circle.

## 2. METHODS

Fig.1 gives simplified overviews of UNet, UNet++ and the proposed UNet 3+. Compared with UNet and UNet++, UNet 3+ combines the multi-scale features by re-designing skip connections as well as utilizing a full-scale deep supervision, which provides fewer parameters but yields a more accurate position-aware and boundary-enhanced segmentation map.

### 2.1. Full-scale Skip Connections

The proposed full-scale skip connections convert the interconnection between the encoder and decoder as well as intra-connection between the decoder sub-networks. Both UNet with plain connections and UNet++ with nested and dense connections are short of exploring sufficient information from full scales, failing to explicitly learn position and boundary of an organ. To remedy the defect in UNet and UNet++, each decoder layer in UNet 3+ incorporates both smaller- and same-scale feature maps from encoder and larger-scale feature maps from decoder, which capturing fine-grained details and coarse-grained semantics in full scales.

As an example, Fig. 2 illustrates how to construct the feature map of $X_{De}^3$. Similar to the UNet, the feature map from the same-scale encoder layer $X_{En}^3$ are directly received in the decoder. In contrast to the UNet, a set of inter encoder-decode skip connections delivers the low-level detailed information from the smaller-scale encoder layer $X_{En}^1$ and $X_{En}^2$, by applying non-overlapping max pooling operation; while a chain of intra decoder skip connections transmits the high-level semantic information from larger-scale decoder layer $X_{De}^4$ and $X_{De}^5$, by utilizing bilinear interpolation. With the five same-resolution feature maps in hand, we need to further unify the number of channels, as well as reduce the superfluous information. It occurred to us that the convolution with 64 filters of size 3 × 3 can be a satisfying choice. To seamlessly merge the shallow exquisite information with deep semantic information, we further perform a feature aggregation mechanism on the concatenated feature map from five scales, which consists of 320 filters of size 3 × 3, a batch normalization and a ReLU activation function. Formally, we formulate the skip connections as follows: let $i$ indexes the down-sampling layer along the encoder, $N$ refers to the total number of the encoder. The stack of feature maps represented by $X_{De}^i$ is computed as:

$$X_{De}^i = \begin{cases} X_{En}^i, & i = N \\ \mathcal{H}\left(\left[\underbrace{C(\mathcal{D}(X_{En}^k))_{k=1}^{i-1}, C(X_{En}^i)}_{Scales: 1^{th} \sim i^{th}}, \underbrace{C(\mathcal{U}(X_{De}^k))_{k=i+1}^{N}}_{Scales:(i+1)^{th} \sim N^{th}}\right]\right), & i = 1, \cdots, N-1 \end{cases} \quad (1)$$

where function $C(\cdot)$ denotes a convolution operation, $\mathcal{H}(\cdot)$ realizes the feature aggregation mechanism with a convolution followed by a batch normalization and a ReLU activation function. $\mathcal{D}(\cdot)$ and $\mathcal{U}(\cdot)$ indicate up- and down-sampling operation respectively, and $[\cdot]$ represents the concatenation.

It is worth mentioning that our proposed UNet 3+ is more efficient with fewer parameters. In the encoder sub-network, UNet, UNet++, and UNet 3+ share the same structure, where $X_{En}^i$ has $32 \times 2^i$ channels. As for the decoder, the depth of feature map in UNet is symmetric to the encoder, and thus $X_{De}^i$ also has $32 \times 2^i$ channels. The number of parameters in $i^{th}$ decoder stage of UNet ($P_{U-De}^i$) can be computed as:

$$P_{U-De}^i = D_F \times D_F \times \left[d(X_{De}^{i+1}) \times d(X_{De}^i) + d(X_{De}^i)^2 + d(X_{En}^i + X_{De}^i) \times d(X_{De}^i)\right] \quad (2)$$

where $D_F$ is the convolution kernel size, $d(\cdot)$ denotes the depth of the nodes. When it comes to UNet++, it makes use of a dense convolution block along each skip pathway, where $P_{U++-De}^i$ can be computed as:

$$P_{U++-De}^i = D_F \times D_F \times \left[d(X_{De}^{i+1}) \times d(X_{De}^i) + d(X_{De}^i)^2 + d(X_{En}^i + \sum_{k=1}^{N-1-i} X_{Me}^{i,k} + X_{De}^i) \times d(X_{De}^i)\right] \quad (3)$$

As can be seen, $P_{U++-De}^i$ is larger than $P_{U-De}^i$. While in UNet 3+, each decoder feature map is derived from $N$ scales, yielding $64 \times N$ channels. $P_{U3+-De}^i$ can be computed as:

$$P_{U3+-De}^i = D_F \times D_F \times \left[\left(\sum_{k=1}^{i} d(X_{En}^k) + \sum_{k=i+1}^{N} d(X_{De}^k)\right) \times 64 + d(X_{De}^i)^2\right] \quad (4)$$

For the sake of the channel reduction, the parameters in UNet 3+ is fewer than those in UNet and UNet++.

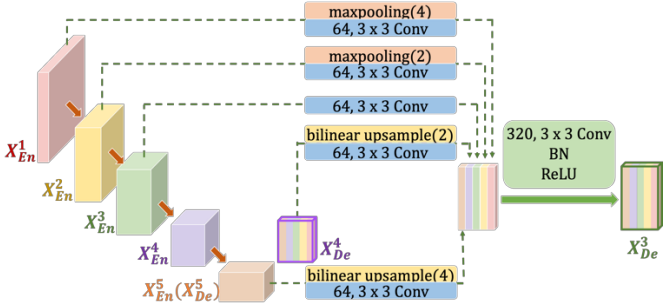

Fig. 2: Illustration of how to construct the full-scale aggregated feature map of third decoder layer $X_{De}^3$.

## 2.2. Full-scale Deep Supervision

In order to learn hierarchical representations from the full-scale aggregated feature maps, the full-scale deep supervision is further adopted in the UNet 3+. Compared with the deep supervision performed on generated full-resolution feature map in UNet++, the proposed UNet 3+ yields a side output from each decoder stage, which is supervised by the ground truth. To realize deep supervision, the last layer of each decoder stage is fed into a plain 3 × 3 convolution layer followed by a bilinear up-sampling and a sigmoid function.

To further enhance the boundary of organs, we propose a multi-scale structural similarity index (MS-SSIM) [9] loss function to assign higher weights to the fuzzy boundary. Benefiting from it, the UNet 3+ will keep eye on fuzzy boundary as the greater the regional distribution difference, the higher the MS-SSIM value. Two corresponding $N \times N$ sized patches are cropped from the segmentation result $P$ and the ground truth mask $G$, which can be denoted as $p = \{p_j : j = 1, \ldots, N^2\}$ and $g = \{g_j : j = 1, \ldots, N^2\}$, respectively. The MS-SSIM loss function of $p$ and $g$ is defined as:

$$\ell_{ms-ssim} = 1 - \prod_{m=1}^{M} \left(\frac{2\mu_p\mu_g + C_1}{\mu_p^2 + \mu_g^2 + C_1}\right)^{\beta_m} \left(\frac{2\sigma_{pg} + C_2}{\sigma_p^2 + \sigma_g^2 + C_2}\right)^{\gamma_m} \quad (5)$$

where $M$ is the total number of the scales, $\mu_p$, $\mu_g$ and $\sigma_p$, $\sigma_g$ are the mean and standard deviations of p and g, $\sigma_{pg}$ denotes their covariance. $\beta_m$ and $\gamma_m$ define the relative importance of the two components in each scale, which are set according to [9]. Two small constants $C_1 = 0.01^2$ and $C_2 = 0.03^2$ are added to avoid the unstable circumstance of dividing by zero. In our experiment, we set the scale to 5 based on [9].

By combining focal loss ($\ell_{fl}$) [10], MS-SSIM loss ($\ell_{ms-ssim}$) and IoU loss ($\ell_{iou}$) [11], we develop a hybrid loss for segmentation in three-level hierarchy – pixel-, patch- and map-level, which is able to capture both large-scale and fine structures with clear boundaries. The hybrid segmentation loss ($\ell_{seg}$) is defined as:

$$\ell_{seg} = \ell_{fl} + \ell_{ms-ssim} + \ell_{iou} \quad (6)$$

## 2.3. Classification-guided Module (CGM)

In the most medical image segmentations, the appearance of false-positives in a non-organ image is an inevitable circums-tance. It is, in all probability, caused by noisy information from background remaining in shallower layer, leading to the phenomenon of over-segmentation. To achieve more accurate segmentation, we attempt to solve this problem by adding an extra classification task, which is designed for predicting the input image whether has organ or not.

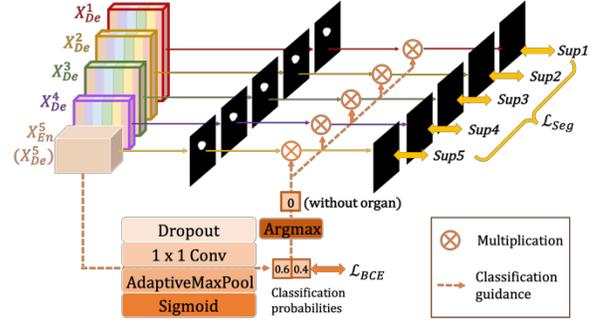

Fig. 3: Illustration of classification-guided module (CGM).

As depicted in Fig. 3, after passing a series of operations including dropout, convolution, maxpooling and sigmoid, a 2-dimensional tensor is produced from the deepest-level $X_{En}^5$, each of which represents the probability of with/without organs. Benefiting from the richest semantic information, the classification result can further guide each segmentation side-output in two steps. First, with the help of the argmax function, 2-dimensional tensor is transferred into a single output of {0,1}, which denotes with/without organs. Subsequently, we multiply the single classification output with the side segmentation output. Due to the simpleness of binary classification task, the module effortlessly achieves accurate classification results under the optimization of Binary Cross Entropy loss function [12], which realizes the guidance for remedying the drawback of over-segmentation on none-organ image.

## 3. EXPERIMENTS AND RESULTS

### 3.1. Datasets and Implementation

The method was validated on two organs: the liver and spleen. The dataset for liver segmentation is obtained from the ISBI LiTS 2017 Challenge. It contains 131 contrast-enhanced 3D abdominal CT scans, of which 103 and 28 volumes are used for training and testing, respectively. The spleen dataset from the hospital passed the ethic approvals, containing 40 and 9 CT volumes for training and testing. In order to speed up training, the input image had three channels, including the slice to be segmented and the upper and lower slices, which was cropped to 320×320. We utilized the stochastic gradient descent to optimize our network and its hyper parameters were set to the default values. Dice coefficient was used as the evaluation metric for each case.

### 3.2. Comparison with UNet and UNet++

In this section, we first compared the proposed UNet 3+ with UNet and UNet++. The loss function used in each method is the focal loss.

Table 1: Comparison of UNet, UNet++, the proposed UNet 3+ without deep supervision (DS) and UNet 3+ on liver and spleen datasets in terms of Dice metrics. The best results are highlighted in **bold**. The loss function used in each method is focal loss.

| Architecture | Vgg-16 | | | ResNet-101 | | | $Dice_{average}$ |
|---|---|---|---|---|---|---|---|
| | Params | $Dice_{liver}$ | $Dice_{spleen}$ | Params | $Dice_{liver}$ | $Dice_{spleen}$ | |
| UNet | 39.39M | 0.9206 | 0.9023 | 55.90M | 0.9387 | 0.9332 | 0.9237 |
| UNet++ | 47.18M | 0.9278 | 0.9230 | 63.76M | 0.9475 | 0.9423 | 0.9352 |
| UNet 3+ w/o DS | **26.97M** | 0.9489 | 0.9437 | **43.55M** | 0.9580 | 0.9539 | 0.9511 |
| UNet 3+ | **26.97M** | **0.9550** | **0.9496** | **43.55M** | **0.9601** | **0.9560** | **0.9552** |

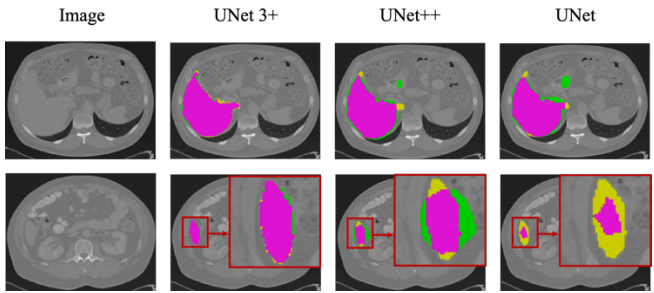

Fig. 4: Qualitative comparisons of ResNet-101-based UNet, UNet++, and proposed UNet 3+ on liver dataset. **Purple areas**: true positive (TP); **Yellow areas**: false negative (FN); **Green areas**: the false positive (FP).

(i) ***Quantitative comparison***: Based on the backbone of Vgg-16 and ResNet-101, Table 1 compares UNet, UNet++ and the proposed UNet 3+ architecture in terms of the number of parameters and segmentation accuracy on both liver and spleen datasets. As seen, UNet 3+ without deep supervision achieves a surpassing performance over UNet and UNet++, obtaining average improvement of 2.7 and 1.6 point between two backbones performed on two datasets. Taking account of both liver and spleen appearing at varying scales in CT slices, UNet 3+ combined with full-scale deep supervision further improved 0.4 point.

(ii) ***Qualitative comparison***: Figure 2 exhibits the segmentation results of ResNet-101-based UNet, UNet++ and UNet 3+ with full-scale deep supervision on liver datasets. It can be observed that our proposed method not only accurately localizes organs but also produces coherent boundaries, even in small object circumstances.

### 3.3. Comparison with the State of the Art

We quantitatively compare our ResNet-101-based UNet 3+ with several recent state-of-the-art methods: PSPNet [3], DeepLabV2 [4], DeepLabV3 [5], DeepLabV3+ [6] and Attention UNet [8]. It is worth mentioning that all results are directly from single-model test without relying on any post-processing tools. Moreover, all networks were optimized by the loss function proposed in their own paper.

Table 2 summarizes the quantitative comparison results. As seen, the proposed hybrid loss function greatly improves the performance by taking pixel-, patch-, map-level optimization in to consideration. Especially, the patch-level MS-SSIM loss function contributes to assign higher weights to the fuzzy boundary, yielding more enhance boundary-aware segmentation map. Moreover, taking advantages of the classification-guidance module, UNet 3+ skillfully avoids the over-segmentation in complex background. As it can be seen, this approach is outstanding compared to all other previous approaches. It is also worth noting that the proposed method outperforms the second best result on the liver (0.9675 against 0.9341) and spleen (0.9620 against 0.9324).

Table 2: Comparison of UNet 3+ and other 5 state-of-the-art methods. The best results are highlighted in **bold**.

| Method | $Dice_{liver}$ | $Dice_{spleen}$ |
|---|---|---|
| PSPNet [3] | 0.9242 | 0.9240 |
| DeepLabV2 [4] | 0.9021 | 0.9097 |
| DeepLabV3 [5] | 0.9217 | 0.9217 |
| DeepLabV3+ [6] | 0.9186 | 0.9290 |
| Attention UNet [8] | 0.9341 | 0.9324 |
| **UNet 3+ (focal loss)** | 0.9601 | 0.9560 |
| **UNet 3+ (Hybrid loss)** | 0.9643 | 0.9588 |
| **UNet 3+ (Hybrid loss + CGM)** | **0.9675** | **0.9620** |

### 4. CONCLUSIONS

In this paper, we proposed a full-scale connected UNet, named UNet 3+ with deep supervision in order to make maximum use of feature maps in full scales for accurate segmentation and efficient network architecture with fewer parameters. The classification-guided module and a hybrid loss function is further introduced to yielding more accurate position-aware and boundary-aware segmentation map. Experimental results on both liver and spleen datasets demonstrated that UNet 3+ surpasses all previous state-of-the-art approaches, highlighting the organs and producing coherent boundaries.

### 5. ACKNOWLEDGMENTS

This work was supported in part by Major Scientific Research Project of Zhejiang Lab under the Grant No.2018DG0ZX01, in part by the Key Science and Technology Innovation Support Program of Hangzhou under the Grant No.20172011A038, and in part by the Grant-in Aid for Scientific Research from the Japanese Ministry for Education, Science, Culture and Sports (MEXT) under the Grant No. 18H03267 and No.17H00754.